\documentclass[11pt]{article}
\pdfoutput = 1
\usepackage[utf8]{inputenc}

\usepackage[top=1in,bottom=1in,left=1in,right=1in]{geometry}
\makeatletter \g@addto@macro\@floatboxreset\centering \makeatother
\usepackage{color}
\definecolor{darkgreen}{rgb}{0,0.5,0}
\definecolor{darkblue}{rgb}{0,0,0.6}
\definecolor{purple}{rgb}{0.4,.2,0.7}
\usepackage[noEucal]{main}
\usepackage{aas_macros,empheq,fancybox,graphicx,multirow}
\usepackage{aas_macros}
\usepackage{dsfont}
\usepackage{esint}
\usepackage{float}
\usepackage{appendix}
\usepackage{dsfont}
\usepackage{stmaryrd}
\usepackage{changepage}
\usepackage{mathabx}
\usepackage{hhline}
\usepackage{booktabs}
\usepackage{tensor} 
\usepackage{footmisc}
\usepackage{tabularray}
\usepackage{float}
\usepackage{caption}
\usepackage{relsize}

\DeclareSymbolFont{cmlargesymbols}{OMX}{cmex}{m}{n}
\let\sumop\relax
\DeclareMathSymbol{\sumop}{\mathop}{cmlargesymbols}{"50}

\usepackage[framemethod=TikZ]{mdframed}

\makeatletter
\DeclareFontFamily{OMX}{MnSymbolE}{}
\DeclareSymbolFont{MnLargeSymbols}{OMX}{MnSymbolE}{m}{n}
\SetSymbolFont{MnLargeSymbols}{bold}{OMX}{MnSymbolE}{b}{n}
\DeclareFontShape{OMX}{MnSymbolE}{m}{n}{
    <-6>  MnSymbolE5
   <6-7>  MnSymbolE6
   <7-8>  MnSymbolE7
   <8-9>  MnSymbolE8
   <9-10> MnSymbolE9
  <10-12> MnSymbolE10
  <12->   MnSymbolE12
}{}
\DeclareFontShape{OMX}{MnSymbolE}{b}{n}{
    <-6>  MnSymbolE-Bold5
   <6-7>  MnSymbolE-Bold6
   <7-8>  MnSymbolE-Bold7
   <8-9>  MnSymbolE-Bold8
   <9-10> MnSymbolE-Bold9
  <10-12> MnSymbolE-Bold10
  <12->   MnSymbolE-Bold12
}{}

\let\llangle\@undefined
\let\rrangle\@undefined
\DeclareMathDelimiter{\llangle}{\mathopen}%
                     {MnLargeSymbols}{'164}{MnLargeSymbols}{'164}
\DeclareMathDelimiter{\rrangle}{\mathclose}%
                     {MnLargeSymbols}{'171}{MnLargeSymbols}{'171}
\makeatother

\DeclareFontFamily{U}{jkpmia}{}
\DeclareFontShape{U}{jkpmia}{m}{it}{<->s*jkpmia}{}
\DeclareFontShape{U}{jkpmia}{bx}{it}{<->s*jkpbmia}{}
\DeclareMathAlphabet{\mathfrakalt}{U}{jkpmia}{m}{it}
\SetMathAlphabet{\mathfrakalt}{bold}{U}{jkpmia}{bx}{it}

\begin{document}

\thispagestyle{empty}

\begin{center}
    ~
    \vskip3cm

     {\LARGE  {\textbf{Challenges to Understanding Celestial Holography \\ \vspace{15pt} from the Bottom Up}}}
    
\vspace{2cm}
Adam Tropper \\
    \vskip1em
    {\it
        Center for the Fundamental Laws of Nature, \\
Harvard University, Cambridge, MA 02138\\ \vskip1mm
         \vskip1mm
    }
    \vskip5mm
    \tt{adam$\_$tropper@g.harvard.edu}
\end{center}
\vspace{10mm}

\thispagestyle{empty}
\begin{abstract}
\noindent 
In the bottom-up approach to celestial holography, it is tempting to define celestial amplitudes by transforming momentum-space amplitudes order by order in perturbation theory. We test this prescription in the exactly solvable two-dimensional Sinh-Gordon model. Because the full non-perturbative $S$-matrix is known, we can compare two operations directly: first transform and then expand, or first expand and then transform. They do not agree, already at leading nontrivial order in the coupling. More broadly, this suggests that naive term-by-term celestial transforms should not be assumed valid in generic quantum field theories with asymptotic weak-coupling expansions. This has an immediate consequence for proposed CCFT duals: if one tries to test them by taking celestial transforms of perturbative bulk amplitudes term-by-term, a mismatch need not falsify the proposal. This makes bottom-up tests of celestial dualities far more subtle than one might have expected.
\end{abstract}
\newpage
\setcounter{page}{1} 
\pagenumbering{arabic}

\setcounter{tocdepth}{2}
{\hypersetup{linkcolor=black}
\fontsize{11.4pt}{11pt}\selectfont
\tableofcontents
}
\section{Motivation}

While there are many well-known examples of AdS/CFT coming from string-theoretic constructions, one can also formulate AdS holography “from the bottom up” \cite{Heemskerk:2009pn, Heemskerk:2010ty, Fitzpatrick:2012cg}. Even partial information about the bulk effective theory already determines nontrivial features of the boundary dual. For example, the masses of light bulk fields tell us the spectrum of low-dimension operators in the CFT. Moreover, certain \textit{symmetries} of the dual CFT can be understood from the bottom up alone --- for example, a bulk graviton/gauge field is dual to a boundary stress tensor/conserved current obeying the appropriate Ward identities. Something similar is true in celestial holography. Knowing the light degrees of freedom in the bulk informs us about corresponding classes of operators in the celestial CFT, while infrared structures such as soft theorems, asymptotic symmetries, collinear limits, and light-ray operators all have natural interpretations as symmetry data of the celestial theory.

Another sense of “bottom-up” holography widely used in the AdS context concerns \textit{dynamics}. Starting from a bulk effective Lagrangian, one may compute boundary correlators and OPE coefficients order by order in perturbation theory using Witten diagrams \cite{Witten:1998qj}. It is tempting to expect an analogous story in celestial holography: namely that perturbative bulk scattering data might similarly determine celestial correlators order by order. It is this sense of bottom-up holography that we will examine in this paper.

When computing scattering amplitudes in momentum-space, we often present the results in a perturbative expansion in some small coupling, $g$. The scattering amplitude takes the form
\begin{equation}
    S(p_1,...,p_n;g) = \sumop_{k=0}^\infty g^k \hspace{2pt}  S_k(p_1,...,p_n) 
    \label{eqn: pert MA}
\end{equation}
The celestial amplitude is defined as the integral transform of the momentum space amplitude with a set of conformal primary wavefunctions $\varphi_{\Delta_i}(z_i,\bar{z}_i;p_i)$ \cite{Pasterski:2016qvg, Pasterski:2017kqt}
\begin{equation}
    \widetilde{S}(\Delta_i,z_i,\bar{z}_i;g) = \int \prod_{i=1}^n \varphi_{\Delta_i}(z_i,\bar{z}_i;p_i) \hspace{2pt} S(p_1,...,p_n;g)
    \label{eqn: mellin transform}
\end{equation}
Such celestial amplitudes are also parameterized by this small coupling constant, so they too may be expanded in $g$. The bottom-up approach to celestial holography suggests that one can compute this series expansion order by order in $g$. To do so, one simply integrates each term in the momentum-space series expansion and matches powers of $g$
\begin{equation}
    \widetilde{S}(\Delta_i,z_i,\bar{z}_i;g) = \sumop_{k=0}^\infty g^k \hspace{2pt} \widetilde{S}_k(\Delta_i,z_i,\bar{z}_i) \overset{\color{red}\textbf{?}\color{black}}{=} \sumop_{k=0}^\infty g^k \hspace{2pt} \int \prod_{i=1}^n \varphi_{\Delta_i}(z_i,\bar{z}_i;p_i) \hspace{2pt} S_k(p_1,...,p_n)
    \label{eqn: CCFT from the bottom up}
\end{equation}
Equation \eqref{eqn: CCFT from the bottom up} tells us that if we know the momentum space $S$-matrix to order $g^k$ and manage to compute the integrals, then we have understood what the celestial $S$-matrix looks like to order $g^k$ as well. Such knowledge \textit{implicitly defines} the celestial CFT to arbitrary order in $g$.

The problem is that this prescription quietly assumes that the perturbative sum and the celestial transform \textit{commute}. In both ordinary QFT and string theory, the perturbative expansion has zero radius of convergence! This is because, as we continue further in the perturbative expansion, there are factorially many Feynman diagrams which contribute to a certain scattering process, so the coefficient in front of $g^k$ gets arbitrarily large \cite{Dyson, Lipatov, Zinn-Justin, Grassi:2014cla, Shenker:1990uf}. Is such a term-by-term definition of celestial amplitudes still valid?

In this paper, we test this question using the exactly solvable $2d$ QFT. The exact non-perturbative $S$-matrix is known, so we can compute the celestial amplitude in two ways: first transform the exact answer and then expand in $g$, or first expand the momentum-space amplitude and then transform term by term. The two answers
\textit{disagree} already at leading nontrivial order. Thus, the naive bottom-up prescription fails in the simplest setting where it can be checked exactly.

\section{Case Study: The Sinh-Gordon Model}
\label{sec: sinh-gordon}

To test the bottom-up prescription, we turn to the Sinh-Gordon model: an exactly solvable, two-dimensional integrable quantum field theory with a known non-perturbative S-matrix.\footnote{We focus on 2d quantum field theory for pragmatic reasons: no non-perturbative $S$-matrices for interacting theories in higher dimensions have been constructed. As we argue in Section \ref{sec: beyond toy models}, the lesson is not special to two dimensions.} The theory is defined by the Lagrangian
\begin{equation}
    \mathcal{L} = \frac{1}{8\pi} (\partial \varphi)^2 + 2 \mu \cosh\big(\sqrt{2} g \hspace{1pt} \varphi\big)
    \label{eqn: sinh-gordon Lagrangian}
\end{equation}
where $g$ is a dimensionless coupling constant and $\mu$ sets a mass scale. The theory has a single species of asymptotic particle with mass \cite{Zamolodchikov:1995xk}
\begin{equation}
    m = \frac{8 \sqrt{\pi} \hspace{2pt} (1+g^2)}{ \Gamma\big(\frac{1}{2+2g^2}\big)\Gamma\big(\frac{g^2}{2+2g^2}\big)} \hspace{2pt} \bigg(\pi \mu\hspace{2pt} g^{-4g^2}\hspace{2pt}\frac{\Gamma(g^2)}{\Gamma(1-g^2)}\bigg)^{1/2(1+g^2)}
    \label{eqn: particle mass}
\end{equation}

In $2d$, we parameterize the momentum of a massive particle in terms of the rapidity $\theta$ according to $p^\mu_i = m_i(\cosh \theta_i,\sinh\theta_i)$. The $2 \rightarrow 2$ $S$-matrix for the Sinh-Gordon model is

\begin{equation}
    \textbf{S}(\theta_1,\theta_2,\theta_1',\theta_2';g) = S(\theta_1-\theta_2;g) \hspace{2pt} \delta(\theta_1 - \theta_1')\delta(\theta_2 - \theta_2')
\end{equation}
where primed vs unprimed quantities distinguish between outgoing and incoming particles. The delta functions play a role analogous to momentum conservation insofar is they demand that the rapidities of individual particles are conserved during the scattering process. The function $S(\theta;g)$ is the all-important function controlling the scattering \cite{Zamolodchikov:1978xm}
\begin{equation}
    S(\theta;g) = \frac{\sinh\theta - i \sin \frac{\pi g^2}{1+g^2}}{\sinh\theta + i \sin\frac{\pi g^2}{1+g^2}}
    \label{eqn: momentum space S matrix}
\end{equation}
This $S$ matrix obeys all the standard axioms of local quantum field theory (for a review, see \cite{Dorey:1996gd}) and enjoys an S-duality $g \rightarrow 1/g$ accompanied by sending send $\mu$ to a suitably modified $\widetilde{\mu}$ \cite{Zamolodchikov:1995xk}.

In two bulk dimensions, the celestial sphere degenerates to a point. Celestial amplitudes therefore have no position dependence; they are labeled only by boost weights. The celestial transform is a Fourier transform in rapidity rather than a Mellin transform in energy \cite{Kapec:2022xjw, Duary:2022onm}. For the two-particle amplitude,
\begin{equation}
    \widetilde{\textbf{S}}(\Delta_1,\Delta_2,\Delta_1',\Delta_2';g) = 2\pi \hspace{2pt} \widetilde{S}(\Delta_1+\Delta_1';g)\hspace{2pt}\delta(\Delta_1+ \Delta_2 + \Delta_1' + \Delta_2') 
\end{equation}
where the delta function constraining boost weight is due to bulk Lorentz invariance. Here $\widetilde{S}(\Delta;b)$, which we shall colloquially call the \textit{$2 \rightarrow 2$ celestial S-matrix}, takes the form
\begin{equation}
    \text{Celestial \textit{S}-matrix:} \hspace{50pt}\color{white}\Bigg|\color{black}\widetilde{S}(\Delta;g) = \int d\theta \hspace{2pt} S(\theta;g) \hspace{2pt} e^{i\Delta \theta}
\end{equation}
\vspace{-20pt}

\subsection{The Non-perturbative Celestial $S$-Matrix}

Computing the Fourier transform defining the exact celestial $S$-matrix is a simple exercise following the earlier logic of \cite{Kapec:2022xjw}.\footnote{See also \cite{Duary:2022onm,Stolbova:2023cof,Stolbova:2023smk,Wang:2025quh} for similar studies of celestial $S$-matrices in integrable quantum field theories (though they all exclusively define celestial amplitudes from the ``bottom up'').} We begin by noting that $S(\theta;g)$ is meromorphic in $\theta$ with simple poles at $\theta = i \pi \frac{2 + g^2}{1+g^2} + 2\pi i \hspace{1pt} k$ and $\theta = i \pi \frac{1 + 2g^2}{1+g^2} + 2\pi i \hspace{1pt} k$ for $k \in \mathbb{Z}$. Such poles indicate the presence of resonances. We evaluate the Fourier transform via integration around the following contour

\begin{figure}[H]
    \centering
    \includegraphics[width=0.8\linewidth]{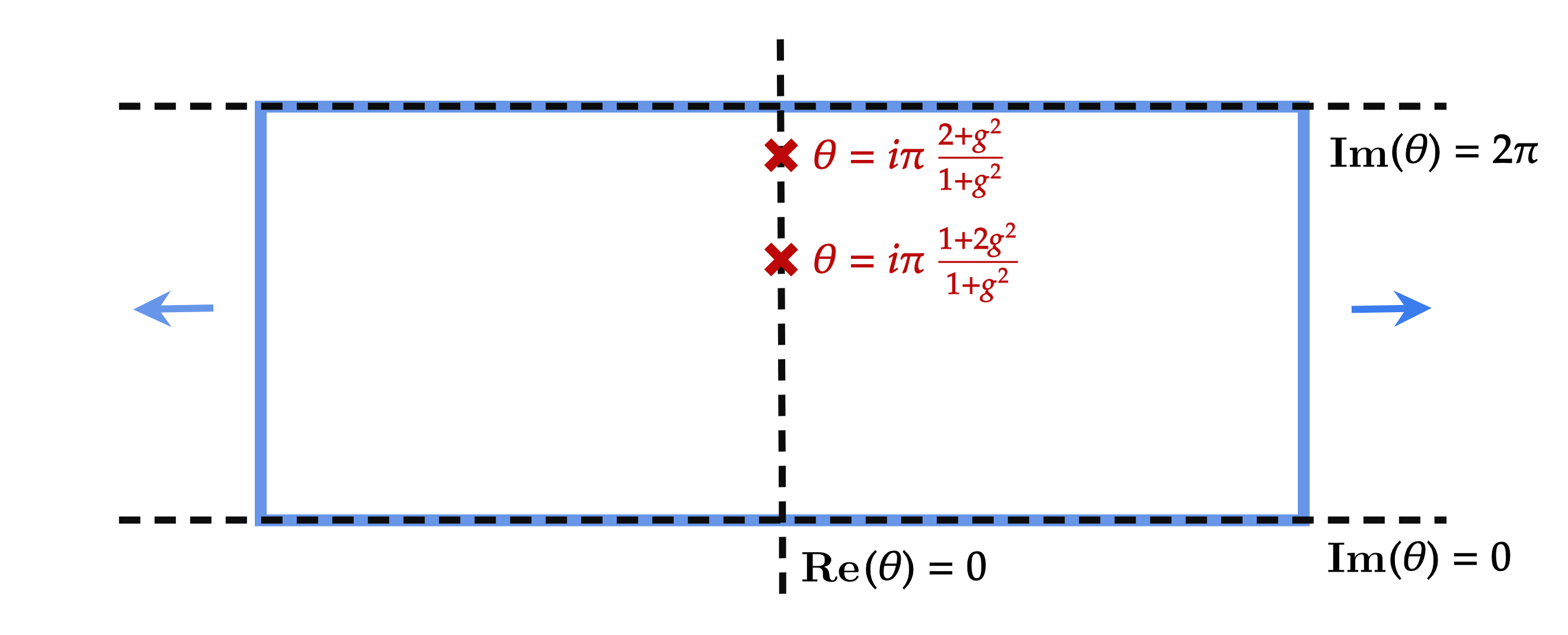}
\end{figure}
\vspace{-10pt}
The $S$-matrix is periodic under $\theta \rightarrow \theta + 2\pi i$ which relates the top and bottom sections of the contour; however, it fails to decay as $\theta \rightarrow \pm \infty$, so the vertical contributions must be treated carefully. A useful trick is leveraging the linearity of the Fourier transform on the space of tempered distributions to write
\begin{equation}
    \widetilde{S}(\Delta;g) = \int_{-\infty}^\infty d\theta \hspace{2pt} e^{i\Delta \theta} +\int_{-\infty}^\infty d\theta\hspace{2pt}\Big[S(\theta;g) - 1\Big] e^{i\Delta \theta} = 2\pi \hspace{1pt} \delta(\Delta) + \int_{-\infty}^\infty d\theta\hspace{2pt}\Big[S(\theta;g) - 1\Big] e^{i\Delta \theta}
\end{equation}
As $\theta \rightarrow \pm \infty$, the second integral behaves nicely and can be evaluated via residues. The result is
\begin{equation}
    \widetilde{S}(\Delta;g) = 2\pi \hspace{1pt}\delta(\Delta) + 2\pi\hspace{2pt} \text{csch}(\pi \Delta) \tan\bigg(\pi \frac{g^2}{1+g^2}\bigg) \bigg[e^{-\pi \Delta/(1+g^2)} - e^{-\pi \Delta g^2/(1+g^2)}\bigg]
    \label{eqn: non-perturbative celestial S-matrix}
\end{equation}
Expanding at small $g$, we find
\begin{equation}
    \widetilde{S}(\Delta;g) = 2\pi \delta(\Delta) + 2\pi^2 \bigg(\tanh \bigg(\frac{\pi\Delta}{2}\bigg) - 1\bigg) g^2 + \mathcal{O}(g^4)
   \label{eqn: series expansion non-perturbative}
\end{equation}
This is the non-perturbative celestial $S$-matrix to leading order in $g$.

\subsection{The Perturbative Celestial $S$-Matrix}

Now reverse the order of operations. The weak-coupling limit $g \rightarrow 0$ is a perturbative limit of the Lagrangian \eqref{eqn: sinh-gordon Lagrangian}, and Feynman diagrams reproduce the small-$g$ expansion of the exact rapidity-space $S$-matrix
\begin{equation}
    S(\theta;g) = \sumop_{k=0}^\infty g^{k} \hspace{2pt} S^{(\text{pert.})}_k(\theta)
\end{equation}
The first non-trivial term in the series expansion is $S^{(\text{pert.})}_2(\theta) = -2 \pi i \hspace{2pt} \text{csch}(\theta).$ Fourier transforming this term gives $\widetilde{S}^{(\text{pert.})}_2(\Delta) = 2\pi^2 \hspace{2pt} \tanh \hspace{-1pt}\left(\frac{\pi \Delta }{2}\right)$ where we have regulated the $\theta = 0$ pole with a principal value prescription. We could have instead regulated with some $i \varepsilon$ prescription, which would have led to a different answer. This choice is immaterial to the conclusion of this article, but we will return to this detail in Section \ref{sec: beyond toy models}. The Fourier expansion of the subleading $\mathcal{O}(g^k)$ terms (including different regularization prescriptions for the integration contour) have been discussed in \cite{Duary:2022onm, Stolbova:2023smk}.

We define the \textit{perturbative celestial amplitude} by Fourier transforming the momentum-space expansion term by term:
\begin{equation}
    \widetilde{S}^{(\text{pert.})}(\Delta;g) = \sumop_{k=0}^\infty g^k \hspace{2pt} \widetilde{S}^{(\text{pert.})}_k(\Delta) \hspace{40pt} \text{where}\hspace{40pt} \widetilde{S}^{(\text{pert.})}_k(\Delta) = \int d \theta \hspace{2pt} e^{i\Delta \theta} \hspace{2pt} S^{(\text{pert.})}_k(\theta)
\end{equation}
At leading, non-trivial order
\begin{equation}
    \widetilde{S}^{(\text{pert.})}(\Delta;g) = 2\pi \delta(\Delta) + 2 \pi^2 \hspace{2pt} \tanh \left(\frac{\pi 
   \Delta }{2}\right) g^2 + \mathcal{O}(g^4)
   \label{eqn: perturbative celestial amplitude}
\end{equation}

\subsection{Challenges to Understanding Celestial Holography From the Bottom Up}

So far, we've extracted the small-$g$ behavior of the celestial $S$-matrix in two ways. In Equation \eqref{eqn: series expansion non-perturbative}, we expanded the exact, non-perturbative result. In Equation \eqref{eqn: perturbative celestial amplitude}, we expanded the momentum-space $S$-matrix first and then Fourier-transformed term by term. \textit{The two results don’t match.} Even the coefficients at $\mathcal{O}(g^2)$ differ. The Sinh-Gordon example shows that even when the exact momentum-space $S$-matrix has a perfectly good weak-coupling expansion, the celestial transform can be sensitive to data lost by the perturbative expansion.

The implication is that bottom-up reconstruction of the celestial dual to quantum gravity is considerably more subtle than one might have expected.\footnote{This should be distinguished from the symmetry-based version of bottom-up celestial holography. The relation between asymptotic symmetries and soft theorems appears to be on much firmer footing: leading soft photon and graviton theorems are understood as exact Ward identities, the associated charges can be constructed directly from the asymptotic/covariant phase space, some loop-corrected soft theorems are also under control, and the connection is also understood from the twistorial viewpoint and in the language of light-ray operators. The obstruction discussed here is instead to the naive perturbative reconstruction of celestial correlators.} Knowledge of perturbative momentum-space amplitudes alone do not, in general, determine the corresponding CCFT correlators, and may even lead to qualitatively incorrect conclusions about their functional form.\footnote{Of course, one can still \textit{define} a ``perturbative'' celestial CFT by applying the transform order by order in the coupling. Such a theory will still have conformal symmetry, and it may capture interesting features of bulk scattering amplitudes; however, this is \textit{not} the CFT which lives on the boundary of an asymptotically flat universe. We hope that this article cautions readers about being careful to distinguish between this perturbative celestial CFT and the true celestial CFT of the bulk theory.}

This evokes a larger issue at play too --- if we have a proposed duality, how can we even verify it? Imagine someone proposes a CCFT dual to type IIB string theory compactified on a Calabi–Yau threefold. A natural test might be to compute a few string amplitudes perturbatively in 
$g_s$, Mellin-transform them, and compare with the correlators predicted by the CCFT. If the two sides don't match, one might think the proposal is flawed. But our result shows that such a mismatch doesn't necessarily mean anything at all --- we never had any right to propose this kind of consistency check in the first place because it's illegal to commute the integrals with the sum defining the string expansion \cite{Grassi:2014cla, Shenker:1990uf}.\footnote{By contrast, this issue does not arise in the best-understood top-down examples, such as Burns space holography. There, the boundary theory is defined independently from the outset the D1-brane worldvolume theory in the topological B-model, while the bulk geometry arises from the backreaction of the same branes in a Maldacena-type setup. The comparison is by matching correlators and OPE data directly rather than relying on a naive term-by-term celestial transform of a perturbative bulk expansion \cite{Costello:2022jpg,Costello:2023hmi}.}

\subsection{Extending the Obstruction Beyond Toy Models}
\label{sec: beyond toy models}

One might worry that the mismatch we found is merely a peculiarity of the Sinh-Gordon model rather than a genuine lesson for celestial holography more broadly. Let us address three natural objections in turn.

The first objection is that at finite $g$ we are scattering a massive particle; however at $g = 0$ the particle becomes massless. Perhaps the issue is related to this singular limit. Masslessness is not the culprit, however. The particle is also massless at the self-dual point $g = 1$; yet, this does not lead to a comparable obstruction. 

A second objection is that the perturbative momentum-space $S$-matrix has a pole at $\theta = 0$, which lies directly on the integration contour. Perhaps the mismatch is simply an artifact of defining the Fourier transform in the principal value sense. Indeed, if one deformed the contour slightly above the pole rather than taking the principal value, one would recover the correct $\mathcal{O}(g^2)$ term in the weak-coupling expansion of the exact celestial $S$-matrix. This works because at finite $g$, the exact $S$-matrix has a pole at $-i \pi \frac{g^2}{1+g^2}$ which approaches the contour from \textit{below} as $g \rightarrow 0$. Perhaps if one was more careful about defining the perturbative prescription, one would have obtained the correct answer from the outset? With access only to perturbative information, however, one can not determine whether the contour should pass above the pole, below it, or be defined in the principal value sense. The sign of the required $i \varepsilon$ is non-perturbative data encoded in the
exact pole structure. Indeed, in the attractive sine-Gordon model, the pole in the lowest breather sector approaches the real axis from \textit{above} in the perturbative limit \cite{Dorey:1996gd}, requiring the \textit{opposite prescription} in order to reproduce the exact weak-coupling answer. 

A related final objection is that while the contour prescription is genuinely non-perturbative information, there is \textit{some} contour prescription which rescues this analysis. One might imagine that in a $4d$ QFT which we understand from the bottom up, being able to regulate all the poles/branch cuts/etc. might save the day, so we still might be able to understand everything with minimal additional data. This is also the wrong lesson to learn: a free boson in $2d$ with a ``gravitational dressing'' is an integrable QFT with the exact 2-particle $S$-matrix $S(\theta;g) = \exp(ig \sinh(\theta))$ \cite{Dubovsky:2012wk, Dubovsky:2017cnj}. The exact celestial $S$-matrix is related to a Bessel function $\widetilde{S}(\Delta;g) = 2 e^{-\pi \Delta/2} K_{i\Delta}(g)$ \cite{Kapec:2022xjw}. Now the problem is even more severe because this exact result doesn't even have a series  expansion in $g$ --- the small $g$ behavior of the exact result goes as $g^{\pm i \Delta}$, not $g^k$ for some $k$. On the other hand, the momentum-space $S$-matrix does have a series expansion and there are \textit{no poles} along the contour. 

The issue was never really about poles: poles are one way for term-by-term integration to fail, but they are not the only one. The real lesson is about perturbation theory itself. A term-by-term celestial transform is justified only if the perturbative expansion has enough \textit{uniform} control to be integrated over kinematic space. Generic quantum field theories present an even sharper problem: perturbation theory is typically asymptotic, with zero radius of convergence even at \textit{fixed} generic kinematics. Thus one often lacks not only uniform control over the celestial integration domain, but \textit{pointwise} convergence before the integral is ever performed. One should therefore not expect term-by-term celestial transforms to be justified in general. The Sinh-Gordon model simply makes this failure fully explicit in a setting where the exact answer is known.

\section*{Acknowledgments}

I would like to thank Wei Fan, Ahmed Sheta, Andrew Strominger and Alexandra Margulies for valuable discussions. This article is inspired by similar work with Dan Kapec who I am grateful to have learned the subject from. I am supported by NSF GRFP grant DGE1745303.

\bibliography{bib.bib}
\bibliographystyle{apsrev4-1long}

\end{document}